\documentclass[letter]{aa} % for the letters 
\usepackage{graphicx}
\usepackage{txfonts}
\usepackage{natbib}
\usepackage[hidelinks]{hyperref}
\begin{document}

   \title{Non-thermal electrons from solar nanoflares}
   
   \subtitle{in a 3D radiative MHD simulation}

   \author{H. Bakke
          \and
          L. Frogner
          \and
          B.V. Gudiksen
          }

   \institute{
   Institute of Theoretical Astrophysics, 
   University of Oslo,
   P.O.Box 1029 Blindern, 
   N-0315 Oslo,
   Norway
   \and
   Rosseland Centre for Solar physics(RoCS),
   University of Oslo, 
   P.O.Box 1029 Blindern, 
   N-0315 Oslo,
   Norway
   }

   \date{}

% \abstract{}{}{}{}{} 
% 5 {} token are mandatory
 
  \abstract
   {We introduce a model for including accelerated particles in pure magnetohydrodynamics (MHD) simulations of the solar atmosphere.}
   {We show that the method is viable and produces results that enhance the realism of MHD simulations of the solar atmosphere.}
   {The acceleration of high-energy electrons in solar flares is an accepted fact, but is not included in the most advanced 3D simulations of the solar atmosphere. The effect of the acceleration is not known, and here we introduce a simple method to account for the ability of the accelerated electrons to move energy from the reconnection sites and into the dense transition zone and chromosphere. }
   {The method was only run for a short time and with low reconnection energies, but this showed that the reconnection process itself changes, and that there is a clear effect on the observables at the impact sites of the accelerated electrons. Further work will investigate the effect on the reconnection sites and the impact sites in detail.}
   {}


   \maketitle
%
%________________________________________________________________

\section{Introduction}
Flares in the solar corona have been observed in wavelength bands from the radio to gamma-rays. The energy released by flares spans at least six orders of magnitude, and the number of flares per energy interval seems to follow a power law. The exponent of the power law has been a constant topic of research since the first investigations of flare number distributions were initiated. If the power is lower than $-2,$ the weakest flares will dominate energetically, and since the strongest flares do not provide enough energy to heat the active solar corona, an exponent equal to or lower than $-2$ is needed in order for the solar corona to be heated by flares.

There are several problems in estimating the flare energy from observations because we only observe the thermal response of the plasma to the reconnection. This response is affected by the local mass density, volume of the reconnection event, and the filling factor of the reconnection event. The important value for the exponent has not been observationally determined as yet, but values higher than $-2$ are more prevalent.

The energy released during a solar flare is known to come from the release of magnetic energy. The energy is supplied by the continuous Poynting flux into the corona by a combination of emerging flux and driving of the existing field by horizontal motions that are present in the photosphere and chromosphere. The released energy can take several forms, none of which are exclusive. According to the classical picture of Petschek reconnection \citep{1964NASSP..50..425P}, the energy is distributed into bulk kinetic energy provided by the acceleration of plasma by the highly bent field lines, thermal energy from the dissipation of the electrical current in the central current sheet, and finally as dissipation of shock waves that are excited at the interface between the regions of different magnetic topology. Observations have also identified another sink for the reconnection energy: acceleration of non-thermal particles. Especially after the launch of the Reuven Ramaty High Energy Solar Spectroscopic Imager (RHESSI) \citep{2002SoPh..210....3L}, the evidence for highly accelerated particles in solar flares has been obvious from the high X-ray and even gamma-ray flux from both the impact of the accelerated particles in the dense transition region and chromosphere and the signature from the reconnection site itself. The origin of the X-ray and gamma-ray signals from the reconnection sites is still debated; it might even be due to the reconnection itself or to the effect of colliding counter-streaming particles \citep{1980ApJ...235.1055E}. %These four energy release bins do not only differ in their observational signatures, they are also able to change the reconnection process itself. If the energy goes into thermal energy locally, the gas pressure increases and if the increase is high enough, the width of the current sheet at the center of the reconnection site will widen. If the energy go into either waves or into accelerating particles, these processes are able to transport the energy away leaving the pressure unaffected and the reconnection will possibly be able to take place at a higher rate than if all was transformed directly into heat. 

Magnetohydrodynamics (MHD), both ideal and resistive, is unable to describe the acceleration of energetic particles because it is a fluid description. Reconnection is a non-ideal process, and resistive MHD is only an approximation of the physical environment at a reconnection site. Consequently, we cannot trust MHD simulations of the central current sheet on scales where the fluid description breaks down. The distribution of the released magnetic energy between thermal, wave, and particle acceleration energy reservoirs might not be properly reproduced. It should be noted that both waves and accelerated particles will deposit some if not all of their energy close to the reconnection site, making the first-order approximation of dumping all the energy as thermal energy better than might be assumed. 

%The damping length of an MHD wave, depends heavily on the type of wave and the local environment. We can speculate that very short-wavelength waves will have a short damping length, while longer wavelengths will have an equally longer damping length. The MHD approximation holds for long scales and so MHD might be able to capture most of the long wavelength waves excited by a reconnection event, while the shorter wavelength modes will dissipate over short length scales, and their energy turned into thermal energy. 

Accelerated particles will thermalise their energy when they collide with ambient particles. The mean free path depends on the velocity of the accelerated particles, so that high-energy particles will on average be able to travel farther than low-energy particles. Accelerated particles therefore seem likely to have a greater impact on the reconnection site itself, but also on the locations where the accelerated particles thermalise. For the solar atmosphere this occurs in the transition region and upper chromosphere. For weak flares, \citet{2014Sci...346B.315T} suggested that the imprint of energetic particles might be observable in the solar transition region and upper chromosphere. 

It has been shown that a fraction of the available charged particles in a reconnection region will be accelerated, and that the maximum acceleration corresponds to acceleration through the large-scale electric field along the current sheet, even though it breaks up during the reconnection process \citep{2013ApJ...771...93B}. It is generally assumed that in MHD, the magnetic energy released by reconnection is directly transferred into heat through Joule-heating in the strong current sheet. A more realistic assumption is that a fraction of the magnetic energy that is released in reconnection events is instead used to accelerate electrons.

We present a model for the energy transport by beams of non-thermal electrons in a realistic radiative MHD simulation of the solar atmosphere. The model consists of a number of steps. It is necessary to identify regions where reconnection is taking place. The energy spectrum of the particles must be inferred from the locally released magnetic energy, which in our simulations is at most of the order of $10^{24}$ ergs, corresponding to nanoflare events \citep{1988ApJ...330..474P}. Finally, the particles need to be followed along the magnetic field while they deposit their energy through collisions with the local plasma.

\section{Identification of reconnection sites}
Reconnection at its heart is a topological effect. Identifying purely topological effects in a 3D simulation is extremely cumbersome and would require tracking the magnetic flux on a very fine scale to identify locations where the topology is not conserved. Instead, a less numerically expensive method of identifying sites of magnetic reconnection is needed. Following \citet{2005mrp..book.....B}, topology is conserved when  
\begin{equation}
\vec{B}\times\left(\vec{\nabla}\times\vec{S}\right) = 0
,\end{equation}
where $\vec{B}$ is the magnetic field vector and $\vec{S}$ is the component of the electric field aligned with the magnetic field. Reconnection then clearly occurs when 
\begin{equation} \label{rec_crit}
\vec{K}=\vec{B}\times\left(\vec{\nabla}\times\vec{S}\right) \ne 0
.\end{equation}
This is readily calculable from pure MHD variables. In order to locate reconnection sites, a reconnection factor is defined as $K_\mathrm{rec} = |\vec{K}|$. Hence, the criterion for reconnection is $K_\mathrm{rec} \ne 0$. Owing to the finite precision in the simulations, $K_\mathrm{rec}$ is almost never exactly zero, and a limit is required for the reconnection factor. Our approach considers reconnection to occur in each grid cell where $K_\mathrm{rec}$ exceeds twice the average value of $K_\mathrm{rec}$ in the horizontal layer that the cell resides in.

\section{Accelerated particles}
Observations have shown that the hard X-ray bremsstrahlung spectrum produced by non-thermal electrons at a coronal loop footpoint typically resembles a power law. Because this feature is largely conserved by the bremsstrahlung emission process, we assume that the energy spectrum of the accelerated particles also follows a power law \citep{2014PhDT........88J}. Electrons are far easier to accelerate than ions. In the interest of studying energy transport over long distances, we therefore only consider accelerated electrons. The number of non-thermal electrons with energy $E$ is then assumed to be proportional to $E^{-\delta}$ when $E$ exceeds some lower cut-off $E_\mathrm{min}$. Electrons with lower energies than $E_\mathrm{min}$ are considered part of the thermal Maxwell--Boltzmann distribution.

We estimate the total energy of a non-thermal electron beam by assigning $50\%$ of the local reconnection energy to the beam. This assumption about the efficiency of the acceleration process is necessary to avoid the need for explicitly modeling the underlying mechanism. The power-law index $\delta$ can in principle be inferred from the corresponding power-law index $\gamma$ of the bremsstrahlung spectrum. We treat $\delta$ as a free parameter, but some loose constraints are provided by observations, which typically imply values of $\delta$ ranging from 3 to 7 (e.g. \cite{1984PhDT........33L}). Finally, we estimate the low-energy cut-off $E_\mathrm{min}$ of an electron beam by insisting that the resulting power-law distribution must intersect the local Maxwell--Boltzmann distribution at the energy given by $E_\mathrm{min}$. This leads to a root-finding problem that we solve using the Newton--Raphson method.

The Lorentz force constrains the accelerated electrons to follow the direction of the magnetic field as they travel through the solar atmosphere. We thus determine the trajectories of the electron beams by tracing magnetic field lines from the reconnection sites. This is achieved with an adaptive Runge--Kutta stepping scheme. Computational limitations on spatial resolution mean that we are unable to resolve the small-scale structure of the current sheets where the particles are accelerated. Since we effectively only know the average direction of the magnetic field at the acceleration sites, the particles are not necessarily distributed correctly between the separate magnetic domains surrounding the reconnection sites. Even so, this is a natural fist step, and we expect that it still qualitatively reproduces how the particles transport energy between the atmospheric layers.

As the particles move through the atmosphere, they take part in various interactions involving the ambient plasma, the electromagnetic field, waves, and radiation. The most important mechanism for energy transfer between the particles and the ambient plasma is Coulomb collisions \citep{1971SoPh...18..489B}. Our treatment of this process is based on a Fokker--Planck equation (e.g. \cite{1990ApJ...359..524M}), which we simplify by neglecting the effect of pitch-angle scattering (by assuming the particles to have a trivial distribution for directions of motion). This leads to a simple expression for the energy loss of individual electrons, which we use to establish a relation between initial electron energies and the depths at which the electrons are thermalised.  Our resulting computation of the energy loss of an electron beam corresponds to dumping the initial energies of the individual electrons at their corresponding thermalisation depths. This highly approximate treatment is necessary to enable the simulation of a large number of electron beams (typically in the order of $10^5$) during each time step of the MHD simulation. 

\section{Simulations}
\begin{figure}[!thb]
    \includegraphics[width=0.5\textwidth]{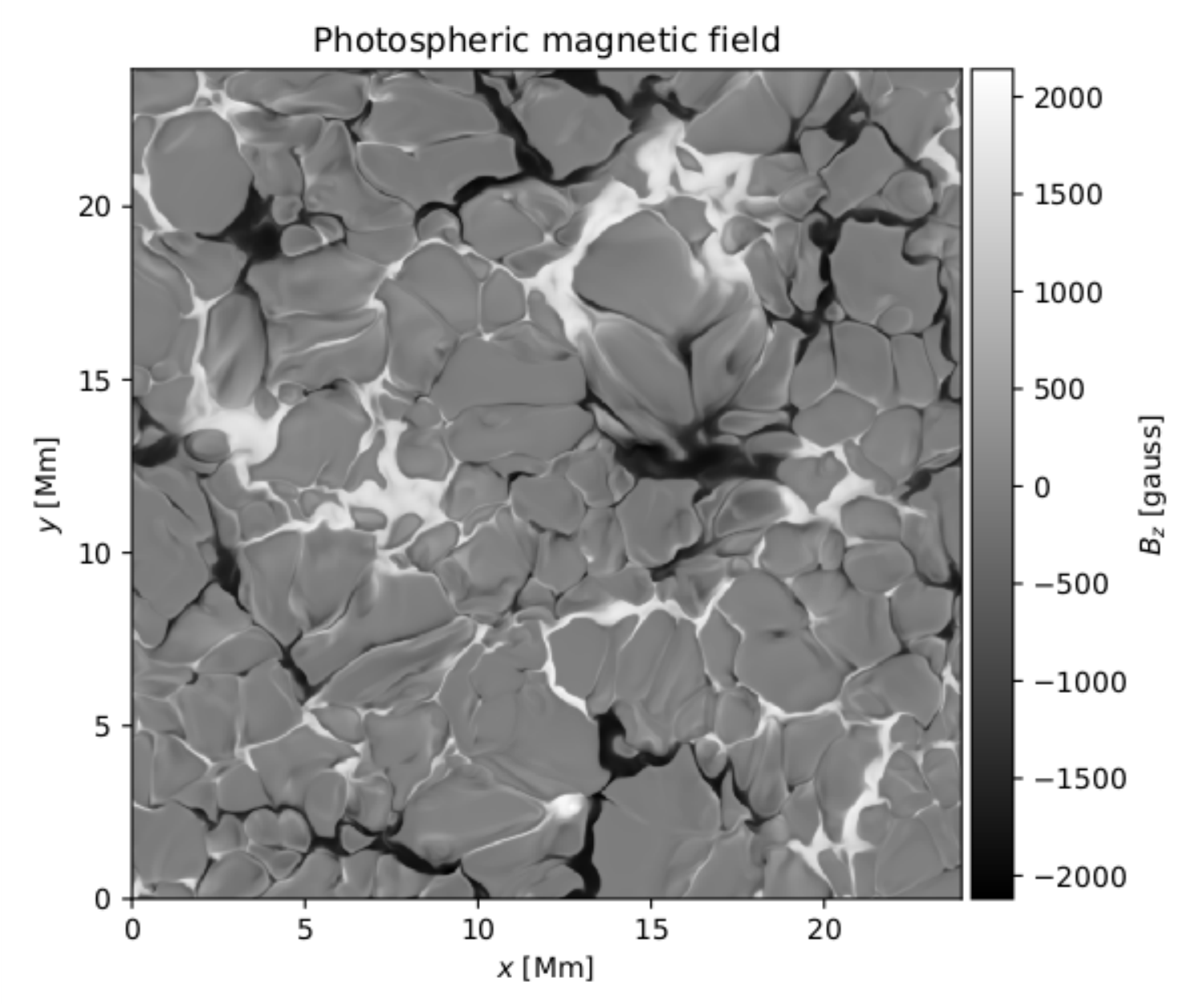}
    \centering
    \caption{Vertical magnetic field component in a horizontal plane lying in the photosphere of the simulation box.}
    \label{fig:bz_photosphere}
\end{figure}

The non-thermal electrons are included in a simulation made with the {\it{\textup{Bifrost}}} code \citep{2011A&A...531A.154G}. The set-up consists of a numerical box encompassing $24\times24$ Mm horizontally and 17 Mm vertically, starting 2.5 Mm below the photosphere. The volume is resolved on a grid of $768^3$ gridpoints, with the vertical $z$-axis having a grid spacing that varies with height. This gives a resolution in the lower atmosphere of 10 km in the region of the photosphere, chromosphere, and transition region that increases to roughly 80 km in the corona. The vertical magnetic field in the photosphere is shown in Fig. \ref{fig:bz_photosphere}. The average unsigned flux is roughly 100 G. The simulation was started with a large amount of horizontal flux being convected through the lower boundary, and at this point in the simulation, most had reached the photosphere, producing a total vertical signed flux very close to zero. The atmosphere was in equilibrium and the corona was heated to roughly 2 MK by the dissipation of the magnetic field. The simulation solved the resistive MHD equations, and included a realistic equation of state and thermal conduction along field lines. No artificial heating terms kept the corona hot; heat was produced by Ohmic dissipation connected to a numerical resistivity, which also worked as a diffusion term in order to keep the code stable. The diffusion term consisted of a small global component as well as a hyper-diffusion component that was able to enhance the diffusion locally. Globally, the magnetic Reynolds number $R_\mathrm{m}$ was typically of order $10^5$, while in regions of strong magnetic gradients, it could become as low as $\sim 50$. From this point, two versions of the simulation were run for roughly 10 solar seconds, one including the non-thermal particles and one without, for reference.

\section{Effect on the corona}
The criterion for reconnection in Equation \eqref{rec_crit} has shown a distribution of reconnection sites that follows loop-like structures, where the coronal loops are anchored to magnetic field footpoints in the simulation photosphere. The indication of magnetic reconnection along coronal loops has also been found by \cite{kanella2017identification}. The inclusion of accelerated particles in the simulation affects the resulting atmosphere, in particular the corona. A direct comparison of two separate simulations, with and without accelerated particles, has indicated a decrease in the number of reconnection sites. This difference indicates a change in the long-term evolution of the corona when electron beams are added to the simulation. The reason for the changes in configuration of the corona is not yet understood, but will be investigated in a follow-up paper. The decreasing number of reconnection sites can be due to different effects. Reconnection occurs as a response to the gradients induced in the field by the injection of Poynting flux. Unless the reconnection event occurs on a very long timescale, the reconnection event will dissipate a finite amount of energy and then cease. When the inclusion of accelerated electrons enhances the reconnection rate by lowering the central gas pressure in the current sheet, the lifetime of the current sheet must consequently be shorter. This would lead to fewer identified reconnection events. Likewise, there might be fewer identified reconnection points because the central current sheet is now narrower as a result of the lower central gas pressure, which would automatically lower the number of grid points that are identified as reconnection sites. These two arguments rely on the fact that the current sheets can be shrunk even further than in the simulation without the accelerated particles. This is not obvious and involves the interplay of several effects, the plasma beta, the diffusion parameters, and so on, and we have not yet investigated this in detail.  The coronal changes can also be due to energy deposited in the lower atmospheric layers by the electron beams, which gives rise to physical processes that alter the corona over time. These processes can be studied through synthetic spectral lines that form in the transition region and chromosphere, where the electrons should deposit most of their energy. The reduction in reconnection sites further implies that the effect of accelerated particles could be required in order to model the corona with high precision. 

\section{Effect on the transition region}
Most of the heating that is due to electron beams occurs in the lower transition region, typically in regions where the magnetic field is relatively strong. This corresponds to coronal loop footpoints, where very many separate beam trajectories converge on the same position.

\begin{figure}[!thb]
    \includegraphics[width=0.5\textwidth]{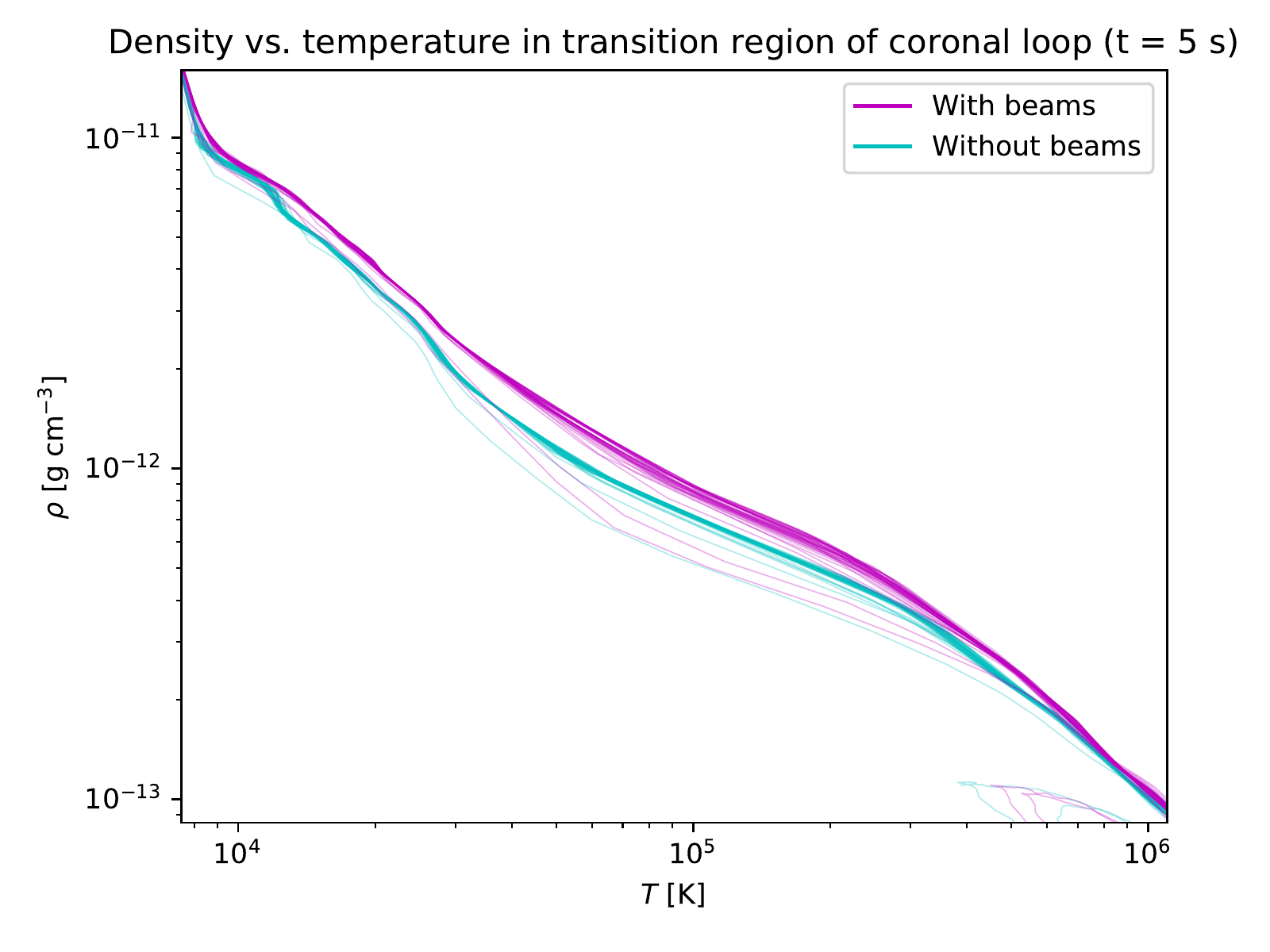}
    \centering
    \caption{Density as a function of temperature along coronal loop field lines passing through an intense transition region beam-heating site. The two sets of field lines correspond to the same coronal loop in two separate simulations, one with electron beams and one without. In both cases the simulation has been evolved for 5 seconds of solar time from the same initial state.}
    \label{fig:coronal_loop_tg_r}
\end{figure}

After about $10$ seconds of beam-heating activity, an average temperature increase in the order of $30\%$ was found at the beam-heating sites. When the beams increase the temperature in the lower transition region, the established equilibrium between heating and radiative cooling is disrupted. This effectively causes the transition region to shift slightly deeper into the atmosphere, where the density is higher. As a consequence, the density of the plasma with a given transition region temperature becomes higher, as shown in Figure \ref{fig:coronal_loop_tg_r}. This suggests that an enhancement of the emission in transition region and chromospheric spectral lines should be produced at the beam-heating sites, which, as we discuss below, is indeed what we found. Qualitatively, the displacement of the transition region to larger depths is an expected response to any form of local energy input \citep{1982ApJ...255..783M}. However, Figure \ref{fig:coronal_loop_tg_r} confirms that the electron beams deposit enough energy for this response to be pronounced.

The temperature increase at the beam-heating sites is accompanied by a corresponding increase in gas pressure. This introduces a pressure gradient between regions of strong and weak beam heating, which leads to an acceleration of plasma away from the regions of most intense beam heating. Typical velocity changes after $10$ seconds in our simulations were in the order of several of km$\;$s$^{-1}$.

\cite{2014Sci...346B.315T} states that numerical models with electron beam heating produce brightening intensities in chromospheric and transition region spectral lines. \ion{Mg}{ii}~h\&k are formed in the chromosphere and provide important diagnostics of this atmospheric layer. Synthetic spectra simulated with the radiative transfer code RH 1.5D \citep{pereira2015rh} have revealed brightening intensities in \ion{Mg}{ii}~h\&k in the simulation with accelerated electrons. The atmospheric response also causes a blue-shift of the lines that most likely results from the increased pressure that causes upflows of hot plasma into the transition region and downflows of cooler plasma into the chromosphere. The signatures in the synthetic spectral lines include line broadening caused by an increase in the vertical velocity. These findings imply that the electrons accelerated by magnetic reconnection deposit their energy in the chromosphere, leaving observable signatures in the synthetic spectral lines. The values for the brightening, the line shifts, and the line broadenings in this simulation are only marginally observable.

\section{Discussion}
When the effect of accelerated particles is included in an MHD simulation of the solar corona, it shows again that the solar atmosphere is a finely balanced system. In the simulation we described, the energy released during a reconnection event is merely moved from the reconnection event itself into the lower transition region and upper chromosphere. In this test, with very low magnetic activity, we already see an effect on the details of the structure of the corona and on the emission that is produced in some of the main observables from the chromosphere, exemplified by the \ion{Mg}{ii}~h\&k lines. The short time-span of the runs prevents us from investigating statistical effects that build up over time. So far, we have not investigated the detailed effect on the current sheets themselves, the cause for the drop in identified grid points with reconnection, or the dependence on the fraction of the released energy that is transferred into accelerated particles. We still argue that the implementation strategy we have chosen is sound.

A key assumption in our model is that the processes governing the transport of energy on the nanoflare level are the same as the processes responsible for ordinary flares, with the distinction only being one of scale. Various statistical studies have suggested that this is reasonable (e.g. \cite{2008ApJ...677.1385C, 1993SoPh..143..275C}).

\begin{figure}[thb]
    \includegraphics[width=0.5\textwidth]{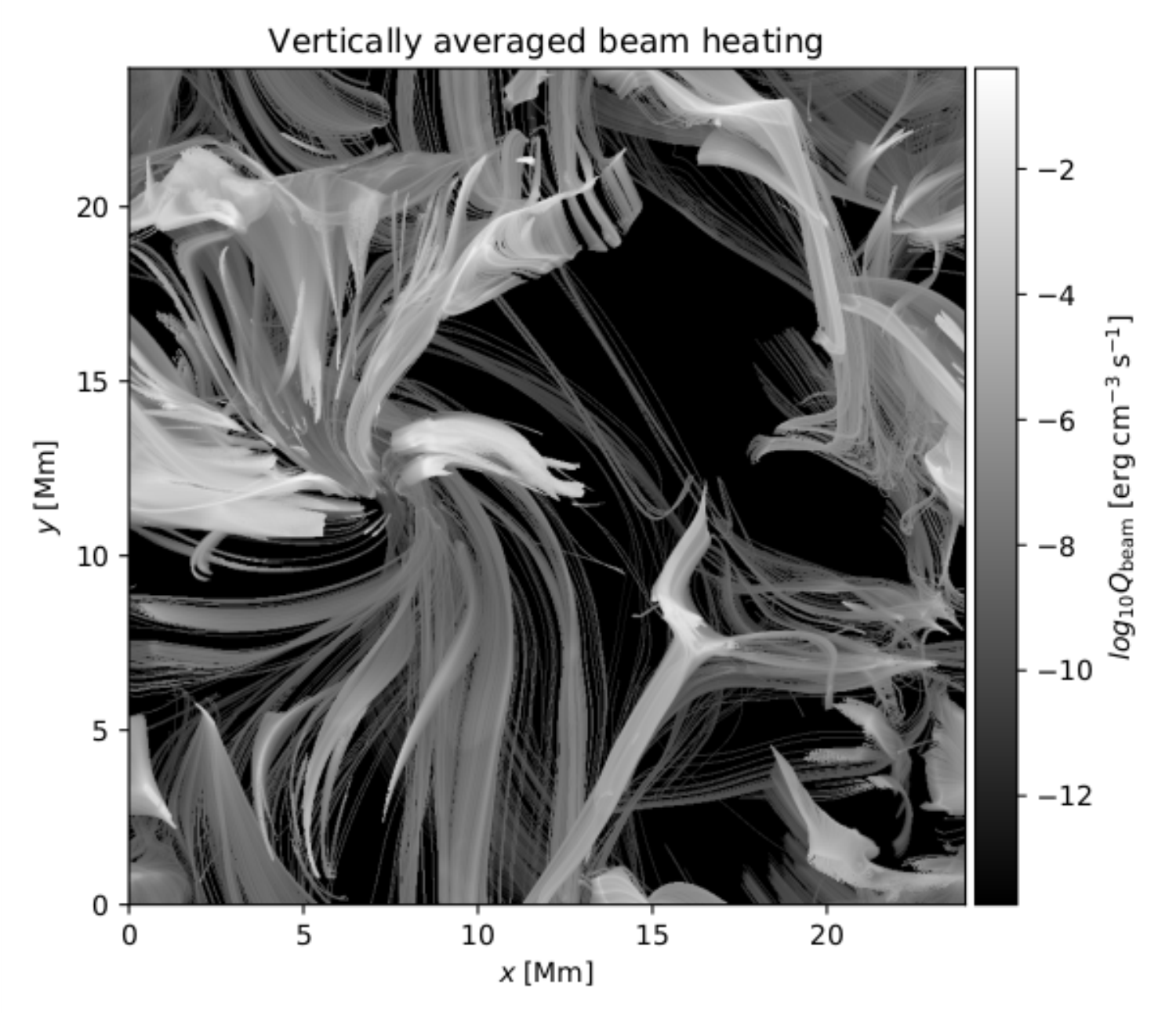}
    \centering
    \caption{Rate of plasma heating that is due to the electron beams in the simulation box, averaged over all depths.}
    \label{fig:qbeam_vertical_avg}
\end{figure}

There are only a few free parameters in this implementation, if the hybrid particle/MHD method employed here is accepted. They are the cut-off value for $K_\mathrm{rec}$, the fraction of the energy released in a reconnection event that is transferred into accelerating electrons, and that the accelerated electrons are distributed in accordance with a power-law distribution in kinetic energy with an index as a free parameter. The chosen cut-off value for $K_\mathrm{rec}$ was based on the computational load of calculating the field lines that the accelerated electrons follow. Figure \ref{fig:qbeam_vertical_avg} shows the energy deposited by electron beams after a single time-step in the simulation. Computing the many extrapolated field lines is  expensive and prohibits the cut-off value for $K_\mathrm{rec}$ from being too low. The fraction of released energy that is transferred into the accelerated electrons was chosen to be 50\% based on \citet{2013ApJ...771...93B}, but this value might change as a function of the local plasma parameters and the details of the reconnection. The power-law distribution of the accelerated electrons is well established even though the power-law index is not. We are not aware of any research that would allow us to pinpoint the correct value of the power-law index, or even if this is variable depending on details of the reconnection process. We therefore chose a value in the middle of the range of published values. 

\begin{acknowledgements}
      This research was supported by the Research Council of Norway through its Centres of Excellence scheme, project number 262622, and through grants of computing time from the Programme for Supercomputing.
\end{acknowledgements}

\bibliographystyle{aa}
\bibliography{electrons.bib}

\begin{thebibliography}{17}
\expandafter\ifx\csname natexlab\endcsname\relax\def\natexlab#1{#1}\fi

\bibitem[{{Baumann} {et~al.}(2013){Baumann}, {Haugb{\o}lle}, \&
  {Nordlund}}]{2013ApJ...771...93B}
{Baumann}, G., {Haugb{\o}lle}, T., \& {Nordlund}, {\r{A}}. 2013, \apj, 771, 93

\bibitem[{{Biskamp}(2005)}]{2005mrp..book.....B}
{Biskamp}, D. 2005, {Magnetic Reconnection in Plasmas}

\bibitem[{{Brown}(1971)}]{1971SoPh...18..489B}
{Brown}, J.~C. 1971, \solphys, 18, 489

\bibitem[{{Christe} {et~al.}(2008){Christe}, {Hannah}, {Krucker}, {McTiernan},
  \& {Lin}}]{2008ApJ...677.1385C}
{Christe}, S., {Hannah}, I.~G., {Krucker}, S., {McTiernan}, J., \& {Lin}, R.~P.
  2008, \apj, 677, 1385

\bibitem[{{Crosby} {et~al.}(1993){Crosby}, {Aschwanden}, \&
  {Dennis}}]{1993SoPh..143..275C}
{Crosby}, N.~B., {Aschwanden}, M.~J., \& {Dennis}, B.~R. 1993, \solphys, 143,
  275

\bibitem[{{Emslie}(1980)}]{1980ApJ...235.1055E}
{Emslie}, A.~G. 1980, \apj, 235, 1055

\bibitem[{{Gudiksen} {et~al.}(2011){Gudiksen}, {Carlsson}, {Hansteen}, {Hayek},
  {Leenaarts}, \& {Mart{\'{\i}}nez-Sykora}}]{2011A&A...531A.154G}
{Gudiksen}, B.~V., {Carlsson}, M., {Hansteen}, V.~H., {et~al.} 2011, \aap, 531,
  A154

\bibitem[{{Jeffrey}(2014)}]{2014PhDT........88J}
{Jeffrey}, N.~L.~S. 2014, PhD thesis, University of Glasgow

\bibitem[{{Kanella} \& {Gudiksen}(2017)}]{kanella2017identification}
{Kanella}, C. \& {Gudiksen}, B.~V. 2017, \aap, 603, A83

\bibitem[{{Leach}(1984)}]{1984PhDT........33L}
{Leach}, J. 1984, PhD thesis, Stanford University

\bibitem[{{Lin} {et~al.}(2002){Lin}, {Dennis}, {Hurford}, {Smith}, {Zehnder},
  {Harvey}, {Curtis}, {Pankow}, {Turin}, {Bester}, {Csillaghy}, {Lewis},
  {Madden}, {van Beek}, {Appleby}, {Raudorf}, {McTiernan}, {Ramaty}, {Schmahl},
  {Schwartz}, {Krucker}, {Abiad}, {Quinn}, {Berg}, {Hashii}, {Sterling},
  {Jackson}, {Pratt}, {Campbell}, {Malone}, {Landis}, {Barrington-Leigh},
  {Slassi-Sennou}, {Cork}, {Clark}, {Amato}, {Orwig}, {Boyle}, {Banks},
  {Shirey}, {Tolbert}, {Zarro}, {Snow}, {Thomsen}, {Henneck}, {McHedlishvili},
  {Ming}, {Fivian}, {Jordan}, {Wanner}, {Crubb}, {Preble}, {Matranga}, {Benz},
  {Hudson}, {Canfield}, {Holman}, {Crannell}, {Kosugi}, {Emslie}, {Vilmer},
  {Brown}, {Johns-Krull}, {Aschwanden}, {Metcalf}, \&
  {Conway}}]{2002SoPh..210....3L}
{Lin}, R.~P., {Dennis}, B.~R., {Hurford}, G.~J., {et~al.} 2002, \solphys, 210,
  3

\bibitem[{{Mariska} {et~al.}(1982){Mariska}, {Doschek}, {Boris}, {Oran}, \&
  {Young}}]{1982ApJ...255..783M}
{Mariska}, J.~T., {Doschek}, G.~A., {Boris}, J.~P., {Oran}, E.~S., \& {Young},
  T.~R., J. 1982, \apj, 255, 783

\bibitem[{{McTiernan} \& {Petrosian}(1990)}]{1990ApJ...359..524M}
{McTiernan}, J.~M. \& {Petrosian}, V. 1990, \apj, 359, 524

\bibitem[{{Parker}(1988)}]{1988ApJ...330..474P}
{Parker}, E.~N. 1988, \apj, 330, 474

\bibitem[{{Pereira} \& {Uitenbroek}(2015)}]{pereira2015rh}
{Pereira}, T. M.~D. \& {Uitenbroek}, H. 2015, \aap, 574

\bibitem[{{Petschek}(1964)}]{1964NASSP..50..425P}
{Petschek}, H.~E. 1964, {Magnetic Field Annihilation}, 425

\bibitem[{{Testa} {et~al.}(2014){Testa}, {De Pontieu}, {Allred}, {Carlsson},
  {Reale}, {Daw}, {Hansteen}, {Martinez- Sykora}, {Liu}, {DeLuca}, {Golub},
  {McKillop}, {Reeves}, {Saar}, {Tian}, {Lemen}, {Title}, {Boerner},
  {Hurlburt}, {Tarbell}, {Wuelser}, {Kleint}, {Kankelborg}, \&
  {Jaeggli}}]{2014Sci...346B.315T}
{Testa}, P., {De Pontieu}, B., {Allred}, J., {et~al.} 2014, Science, 346,
  1255724

\end{thebibliography}
\end{document}